\documentclass[pra,aps,reprint,superscriptaddress,showpacs]{revtex4-1}

\usepackage{graphicx}
\usepackage{verbatim}
\usepackage{comment}
\usepackage{amsmath}
\usepackage{amssymb}
\usepackage{stmaryrd}
\usepackage{color}
\usepackage{xfrac}
\usepackage{units}

\renewcommand{\O}{\mathcal{O}}

\begin{document}

\title{Quantum quenches and many-body localization in the thermodynamic limit}

\author{Baoming Tang} \affiliation{Department of Physics, The
  Pennsylvania State University, University Park, PA 16802, USA}
\affiliation{Department of Physics, Georgetown University, Washington,
  DC 20057, USA} \author{Deepak Iyer} \author{Marcos Rigol}
\affiliation{Department of Physics, The Pennsylvania State University,
  University Park, PA 16802, USA}

\begin{abstract}
  We use thermalization indicators and numerical linked cluster
  expansions to probe the onset of many-body localization in a
  disordered one-dimensional hard-core boson model in the
  thermodynamic limit. We show that after equilibration following 
  a quench from a delocalized state, the momentum distribution 
  indicates a freezing of one-particle correlations at higher values
  than in thermal equilibrium. The position
  of the delocalization to localization transition, identified by the
  breakdown of thermalization with increasing disorder strength, is
  found to be consistent with the value from the level
  statistics obtained via full exact diagonalization of finite chains.
  Our results strongly support the existence of a many-body localized
  phase in the thermodynamic limit.
\end{abstract}


\maketitle

Since the first quantitative discussion of localization by Anderson in
1958 \cite{anderson58}, a large number of experiments have revealed
phenomena governed by localization physics in solid state 
\cite{lee_ramakrishnan_85,kramer93} and atomic 
\cite{Bayfield_1988,Bayfield_1989,Bharucha_1999,Roati_2008} physics.  
In the absence of interactions, destructive interference due to scattering 
off of impurities is responsible for localization \cite{anderson58}. What
happens in the presence of interactions has remained an open problem
whose exploration has become an active area of research over the past
few years.  For weak interactions, perturbative arguments support the
existence of localized phases
\cite{fleishman80,altshuler_gefen_97,gornyi_mirlin_05,basko06}.  For
strong interactions, on the other hand, numerical studies have found
signatures of many-body localization and have explored its implications
\cite{oganesyan_huse_07,znidaric08,pal10,khatami_rigol_12,bardarson_pollmann_12,
 vosk_altman_13,serbyn_papi_13,kjall14,grover14,sirker,vasseur14,altman14,serbyn14}.
Nonetheless, it remains a challenge to conclusively establish that, in
the presence of strong interactions, the delocalization to
localization transition occurs at finite disorder strength in the
thermodynamic limit.

The signatures of localization in experiments are mostly dynamical in
nature, e.g., measurements of the conductivity
\cite{kramer93}. Theoretically, it is difficult to study dynamical
quantities. So, to identify many-body localized phases, it is common to
use the statistics of the energy level spacing instead (see, e.g.,
Refs.~\cite{oganesyan_huse_07,pal10,khatami_rigol_12}). Poissonian level
statistics is expected for localized phases, whereas Wigner-Dyson
statistics is expected for delocalized ones. Equally accessible to
experimental and theoretical studies is a defining, but less
explored, signature of many-body localization---when taken far from
equilibrium, isolated localized systems do not thermalize \cite{nandkishore14}.

Relaxation dynamics and thermalization in isolated many-body quantum
systems is a very active area of current research on its own
\cite{cazalilla_rigol_10,dziarmaga_10,polkovnikov_sengupta_review_11}.
There is growing evidence that generic many-body quantum systems
thermalize after being taken far from equilibrium
\cite{rigol_dunjko_08,rigol_09a,rigol_09b,eckstein_kollar_09,banuls_cirac_11,
rigol_14a}, and that this is a consequence of eigenstate thermalization
\cite{rigol_dunjko_08,rigol_09a,rigol_09b,deutsch_91,srednicki_94,
santos_rigol_10,roux_10,rigol_srednicki_12,neuenhahn_marquardt_12,khatami_pupillo_13,
steinigeweg_herbrych_13,beugeling_moessner_14,sorg_vidmar_14,beugeling_14,kim_14}.
That is, thermalization results from the fact that, for few-body observables, 
individual eigenstates of the Hamiltonian already exhibit thermal properties
\cite{rigol_dunjko_08,deutsch_91,srednicki_94}. This can be pictured as the system 
effectively acting as its own bath. Such a picture breaks down in integrable systems 
\cite{rigol_dunjko_08,rigol_09a,rigol_09b} and in many-body localized ones. 
In the latter, different parts of the system cannot communicate with one another, 
i.e., they cannot be ergodic \cite{nandkishore14}. Numerical calculations in finite 
systems have provided evidence of the breakdown of eigenstate thermalization 
\cite{khatami_rigol_12} and thermalization \cite{khatami_rigol_12,gogolin_muller_11} 
in disordered many-body systems.

Here, we study quantum quenches in disordered isolated systems in the
thermodynamic limit. By a quantum quench it is meant that the initial
state is stationary with respect to an initial Hamiltonian, which is
suddenly changed to a new (time-independent) Hamiltonian. The latter
then drives the (unitary) dynamics of the system. We are interested in
the time average of observables (say, $\hat{O}$) after the
quench. They can be calculated as
$\overline{O(\tau)}=\overline{\text{Tr}[\hat{\rho}(\tau)\hat{O}]}=
\text{Tr}[\overline{\hat{\rho}(\tau)}\hat{O}]\equiv
\text{Tr}[\hat{\rho}_\text{DE}\hat{O}]=O_\text{DE}$, where
$\overline{(\cdot)}=\text{lim}_{\tau'\rightarrow\infty}1/
\tau'\int_0^{\tau'} d\tau\,(\cdot)$ indicates the infinite time
average, $\hat{\rho}(\tau)$ is the density matrix of the time-evolving
state, and $\hat{\rho}_\text{DE}\equiv\overline{\hat{\rho}(\tau)}$ is
the density matrix of the so-called diagonal ensemble (DE)
\cite{rigol_dunjko_08}.  To obtain results in the thermodynamic limit,
we advance a recently introduced numerical linked cluster expansion (NLCE)
for the DE \cite{rigol_14a,wouters_denardis_14,rigol_14b}.
NLCEs for systems in thermal equilibrium were introduced in
Refs.~\cite{rigol_bryant_06,rigol_bryant_07}, and their implementation
was discussed in Ref.~\cite{tang_khatami_13}. When converged,
NLCE calculations provide exact results in the thermodynamic limit.
For quenches in the integrable $XXZ$ chain, this was shown in Refs. [44,45] 
by comparing NLCEs with exact analytic calculations using the Bethe ansatz. 
In this Rapid Communication, thermalization, or the lack thereof, is studied by 
comparing results for observables in the DE and in the grand-canonical 
ensemble (GE).

We focus on a system of impenetrable bosons in one-dimension (1D) with
Hamiltonian $\hat{H}=\hat{H}_{0}+\hat{H}_D$, where
\begin{equation}
  \label{eq:ham-hcb}
  \hat{H}_{0}= \sum_{i}\left[-t(\hat{b}^\dagger_{i}\hat{b}^{}_{i+1}+\text{H.c.}) 
    + V\left(\hat{n}_{i}-\frac{1}{2}\right)\left(\hat{n}_{i+1}-\frac{1}{2}\right)\right]
\end{equation}
is translationally invariant and
$\hat{H}_D=\sum_{i}h_{i}(\hat{n}_{i}-\frac{1}{2})$ is the term with
the disorder. $\hat{b}^\dagger_{i}$ ($\hat{b}_{i}$) creates
(annihilates) a hard-core boson at site $i$ and
$\hat{n}_{i}=\hat{b}^\dagger_{i}\hat{b}_{i}$ is the site number
operator. $t$ stands for the hopping parameter, $V$ for the nearest
neighbor interaction, and $h_{i}$ for the strength of the on-site
disorder. In the spin language, $\hat{H}$ describes a spin-1/2 $XXZ$
model in the presence of a random magnetic field in the
$z$ direction. We select the random field to have a binary
distribution with equal probabilities for $h_{i}=\pm h$. This model
has been recently motivated in the context of ultracold bosons in
optical lattices \cite{sirker}.

We first use full exact diagonalization of finite chains with open
boundary conditions to check whether $\hat{H}$ supports a many-body localized 
phase (as argued in Ref.~\cite{sirker}) and, if it does, the value of the
disorder strength at which such a phase appears. We focus on $V=2t$
(which is the Heisenberg point in the spin model) and set $t=1$ as our
unit of energy. As a first indicator of many-body localization, we
study the averaged ratio of the smaller and the larger of two consecutive 
energy gaps, $r_{n}=\mathrm{min}[\delta^{E}_{n-1},\delta^{E}_{n}]/
\mathrm{max}[\delta^{E}_{n-1},\delta^{E}_{n}]$, where $\delta^{E}_{n}
\equiv E_{n+1}-E_{n}$ is the difference between adjacent energy levels
in the spectrum \cite{oganesyan_huse_07,pal10}. The averaged ratio $r$ is obtained 
by averaging $r_{n}$ over the central half of the spectrum for a given disorder
configuration, and then averaging over disorder configurations.  In
the delocalized phase, one expects $r\approx 0.5359$ and in the
localized one, $r\approx 0.3863$, corresponding to the results for the
Wigner-Dyson and Poissonian distributions \cite{atas_bogomolny_13},
respectively.

\begin{figure}[!t]
  \includegraphics[width=0.95\columnwidth]{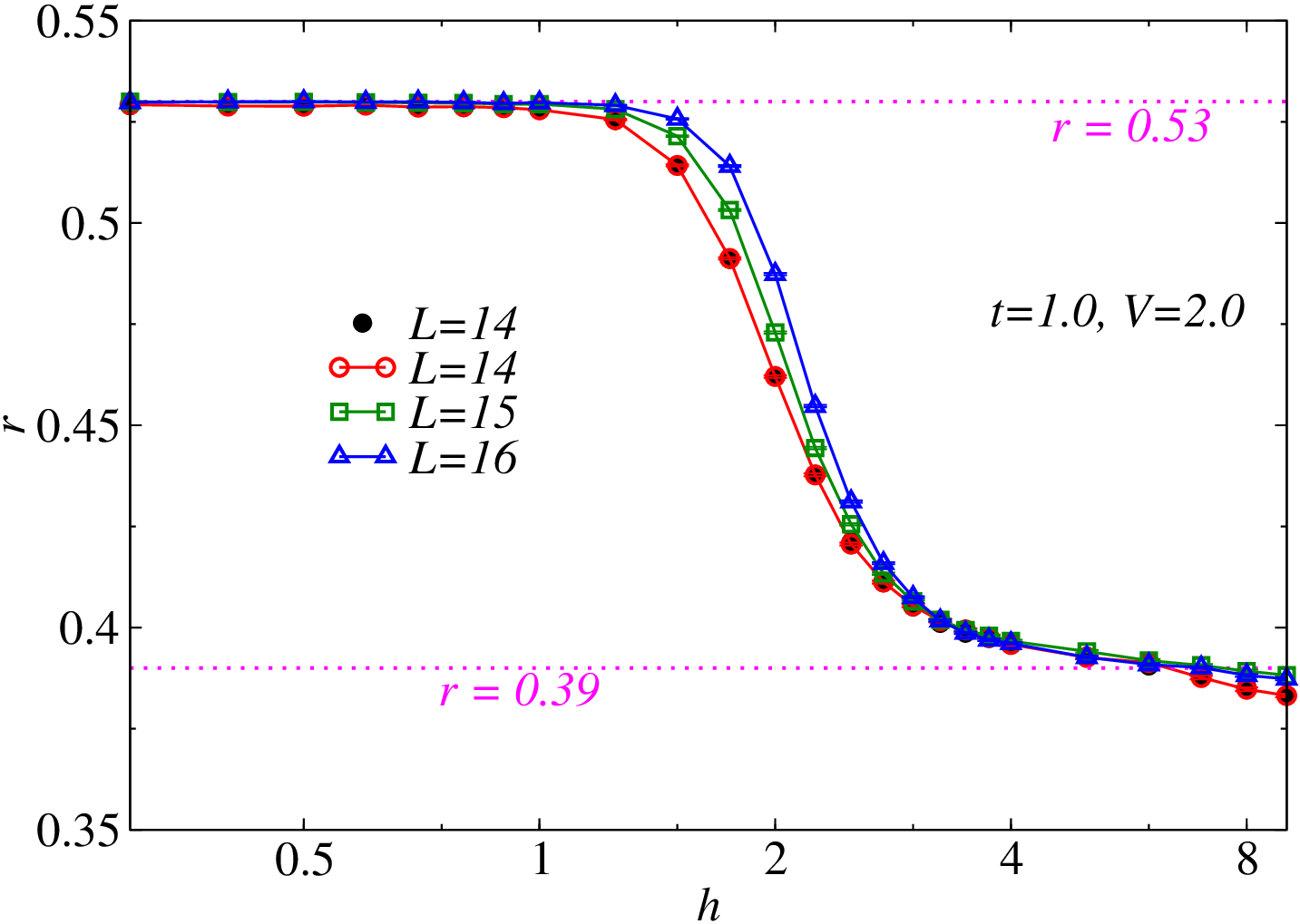}
  \vspace{-0.17cm}
  \caption{(Color online) Exact diagonalization results for 
  the averaged ratio of two consecutive energy gaps $r$ (see text) 
  as a function of disorder strength in chains with 
  $L=14$, $15$, and $16$ sites, and $V=2$. For $L=14$, the energy ratio 
  was computed considering all $2^{14}=16\,384$ disorder field configurations 
  (solid circles). We also show the energy ratio for $L=14$ (open circles), 
  $L=15$ (open squares), and $L=16$ (open triangles) averaging over 9100 random
  samples. The error bars depict one standard deviation. They make apparent 
  that the statistical errors are negligible at the scale of the figure.}
  \label{fig:energyRatioOpen}
\end{figure}

Figure \ref{fig:energyRatioOpen} shows the averaged ratio $r$ as a function of
the strength of the random field $h$ for three system sizes. One can
see that there is a transition from a delocalized to a localized phase
with increasing disorder strength, and that it sharpens with
increasing system size. From the delocalized side, with increasing
$h$, the curves for different system sizes meet in the vicinity of
$h=3.5$, suggesting that the critical $h_c\approx3.5$. Remarkably, for
the same model but with continuous disorder, the transition was found to 
be at around twice this value ($h_c\approx7$) \cite{pal10}.

\begin{figure}[!t]
  \includegraphics[width=0.97\columnwidth]{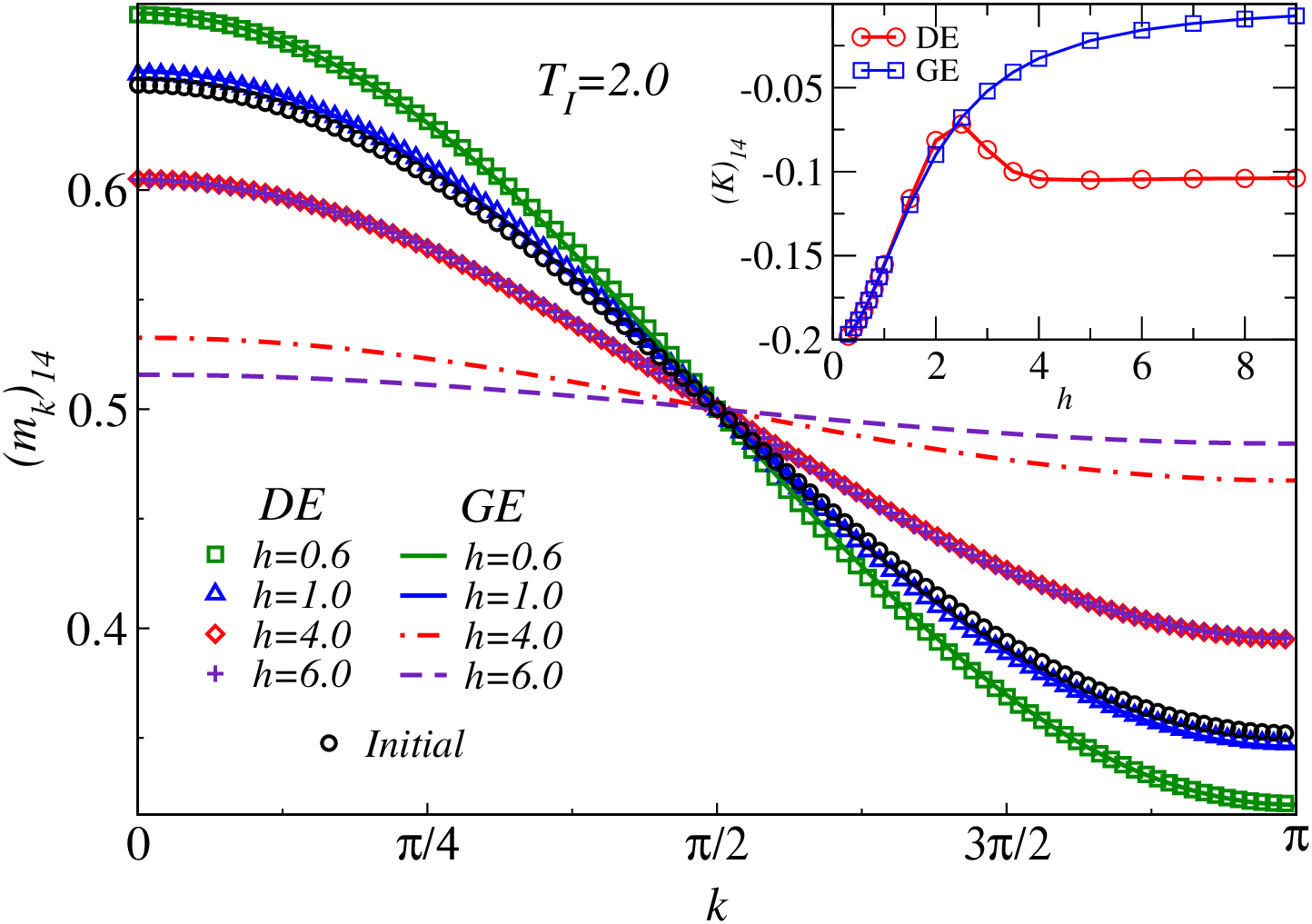}
  \vspace{-0.2cm}
  \caption{(Color online) Last order ($l=14$) of the NLCE calculation
    for the momentum distribution in the initial state with $T_{I}=2$,
    and in the DE and GE after quenches with four different values of
    the disorder strength $h$ (two below and two above the
    delocalization to localization transition). The inset depicts the
    last order of the NLCE for the kinetic energy $K$ after quenches
    as a function of $h$.  Note that, for $h\lesssim2.5<h_c$, the
    results in the DE and the GE are virtually indistinguishable.}
  \label{fig:momentum}
\end{figure}

Now that we have an idea of the disorder strengths that correspond to 
the ergodic and many-body localized phases, we proceed to study quantum
quenches into both regimes. We take the initial state to be in thermal
equilibrium at some temperature $T_I$ for $\hat{H}_I$ with parameters
$t_{I}=0.5$, $V_{I}=2.5$, and $h_{j}=0$ for all $j$, i.e., the initial
state is homogeneous. (We have verified that the results reported are 
robust when changing the initial state, which is, in principle, 
arbitrary.)  After the quench, we take $t=1$, $V=2.0$, and different 
values of $h\neq0$ (as in Fig.~\ref{fig:energyRatioOpen}). 
In all our calculations, the chemical potential $\mu=0$, so that the 
systems are at half filling. NLCEs for the diagonal ensemble allow 
one to compute the infinite-time average of observables in the 
thermodynamic limit for lattice systems evolving unitarily 
\cite{rigol_14a,rigol_14b}. The fundamental NLCE development introduced 
in this Rapid Communication is the ability to deal with systems with disorder.

\begin{figure*}[!t]
  \includegraphics[width=0.9\textwidth]{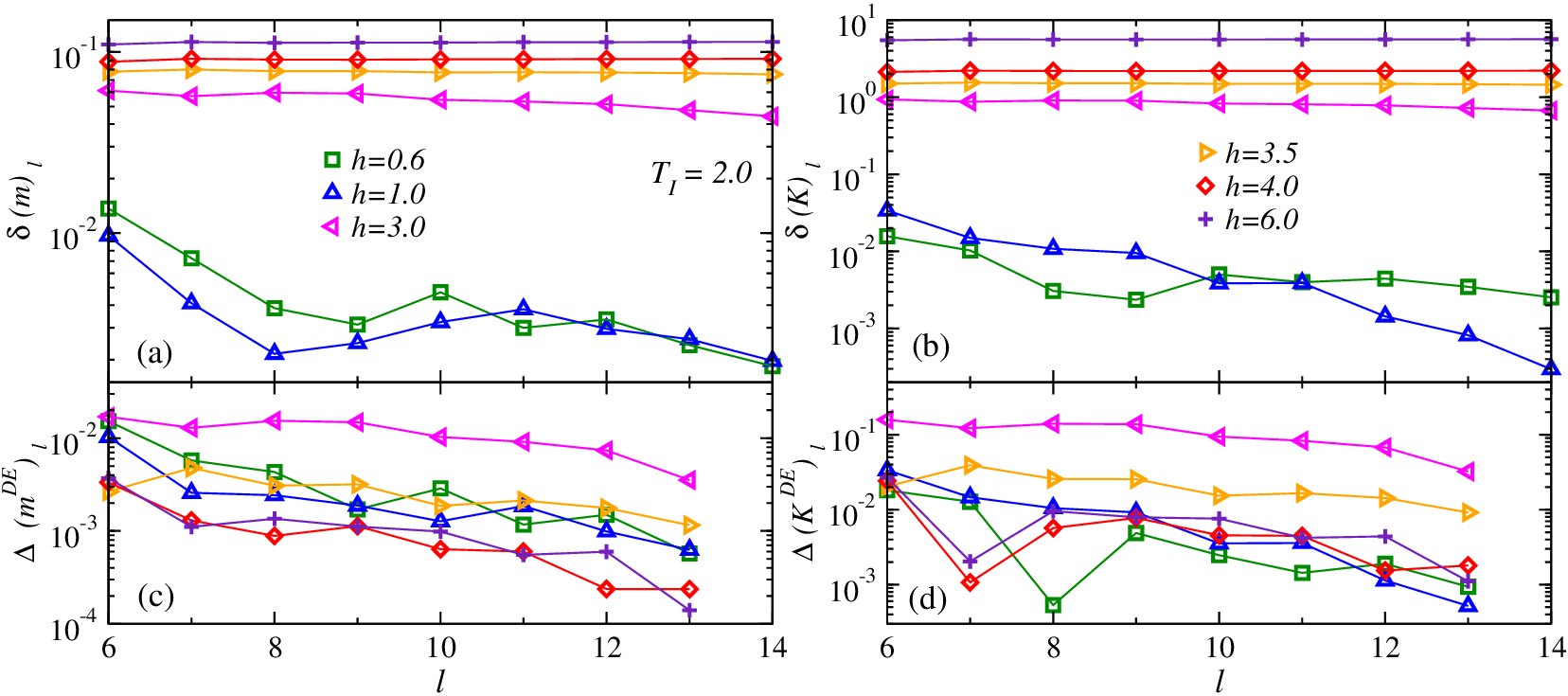}
  \vspace{-0.17cm}
  \caption{(Color online) Relative differences for the momentum
    distribution and the kinetic energy vs $l$ in the NLCE calculation 
    for six values of $h$ and $T_I=2$. (a) $\delta(m)_{l}$, 
    (b) $\delta(K)_{l}$, (c) $\Delta(m^\text{DE})_{l}$, and (d)
    $\Delta(K^\text{DE})_{l}$. Results for $\Delta(m^\text{GE})_{l}$
    and $\Delta(K^\text{GE})_{l}$ are reported in Ref.~\cite{supmat}.
    For $h=4$ and 5, the results for $\delta(m)_{l}$ and $\delta(K)_{l}$
    do not change with changing $l$, i.e., they have converged.}
  \label{fig:deltaKinMom}
\end{figure*}

In translationally invariant systems, NLCEs allow one to calculate the
expectation value of an extensive observable per lattice site in the
thermodynamic limit, $\O$,  as a sum over the contributions from all
clusters $c$ that can be embedded on the infinite lattice:
$\O=\sum_{c}M(c)\times W_{\O}(c)$, where $M(c)$ is the multiplicity of
$c$, defined as the number of ways per site in which cluster $c$ can
be embedded on the lattice. $W_{\O}(c)$ is the weight of $\hat{\O}$ in
cluster $c$, which is calculated recursively using the
inclusion-exclusion principle $W_{\O}(c)=\O(c)-\sum_{s \subset c}
W_{\O}(s)$, where $\O(c)=\textrm{Tr}[\hat{\O}\hat{\rho}_{c}]$ is
computed using full exact diagonalization, with $\hat{\rho}_{c}$ being
the density matrix relevant to the calculation [e.g., of the
grand-canonical ensemble (GE) or the diagonal ensemble (DE)] in
cluster $c$ \cite{rigol_14a,rigol_14b}.

Such an expansion cannot be applied to systems in which translational symmetry 
is broken, e.g., by disorder. However, a disorder average that restores an exact
translational invariance enables once again the use of NLCEs. The two
crucial points that make that possible are: (i) the linear character
of the equations defining the linked cluster expansion, so that 
disorder average can be commuted with the NLCE
summation process, and (ii) the use of binary disorder which, after
averaging over all possible disorder realizations, restores the 
translational symmetry (and also particle-hole symmetry)
of $\hat{H}_0$. Hence, all we need to do in our calculations is
replace $\O(c)=\textrm{Tr}[\hat{\O}\hat{\rho}_{c}]$ for the
translationally invariant case by:
\begin{equation}
  \label{eq:obs_disorder}
  \O(c)=\left\langle \textrm{Tr}[ \hat{\O} \hat{\rho}_{c}]
  \right \rangle_{\rm dis},
\end{equation}
where $\langle \cdot \rangle_{\rm dis}$ represents the disorder
average. Having to compute this additional average reduces our site based
linked cluster expansion from a maximum of 18 sites for translationally 
invariant systems \cite{rigol_14a,wouters_denardis_14,rigol_14b} 
to 14 sites here. We define $\O_{l}^{\mathrm{ens}}$ as the sum over the
contributions of clusters with up to $l$ sites, where
``$\mathrm{ens}$'' could be DE or GE. The temperature used in the GE
calculations to describe the system after the quench is determined 
from a comparison of the energy of DE and the GE by ensuring that
$|E^\text{DE}_{14}-E^\text{GE}_{14}|/|E^\text{DE}_{14}|< 10^{-12}$. We
only report results for values of $T_I$ for which $E^\text{DE}_{14}$
and $E^\text{GE}_{14}$ are converged within machine precision (see
Ref.~\cite{supmat}).

In Fig.~\ref{fig:momentum}, we report the initial momentum distribution 
of a system with $T_I=2$ and the final momentum distribution for different 
values of $h$ after the quench. After the quench, the DE and GE results for 
$h=0.6$ and 1 ($h<h_c$) are indistinguishable from each other, while for 
$h=4$ and 6 ($h>h_c$) they are very different from each other. Remarkably, the 
results that are close to each other for $h>h_c$ are those from the DE. 
The contrast between the DE and GE results in this regime makes apparent 
that there is more coherence in the one-particle sector after 
equilibration than if the system were in thermal equilibrium 
($m^\text{DE}_{k=0}>m^\text{GE}_{k=0}$).  The system
``remembers'' one-particle correlations from the initial state.  This
has also been seen in quasiperiodic systems \cite{gramsch_rigol_12}.
It is easy to understand in the limit of very strong disorder, where
$\hat{H}=\sum_{i}h_{i}(\hat{n}_{i}-\frac{1}{2})$, and, in the
Heisenberg picture, $\hat{b}_i^\dagger(\tau)\hat{b}_j^{}(\tau)=
\exp[i(h_{i}-h_{j})\tau/\hbar]\hat{b}_i^\dagger(0)\hat{b}_j^{}(0)$.
A disorder average over $h_{i}, h_{j}$ (with each being $\pm h$ 
with equal likelihood) reveals that, for a half-filled system,
$m^\text{DE}_{k}=1/4+m_{k}(\tau=0)/2$. Strikingly, a very strong
freezing of correlations is seen right after entering the
many-body localized phase. The results for the kinetic energy,
almost constant in the inset in Fig.~\ref{fig:momentum} for $h>h_c$, 
provide evidence of the robustness of these findings.

To discern which of the differences between the DE and GE seen in
Fig.~\ref{fig:momentum} are due to lack of convergence of the NLCE 
and which are expected to survive in the thermodynamic limit, 
we calculate the following two differences,
\begin{equation}\label{eq:smalldelta}
  \delta(m)_{l} = \frac{\sum_k|(m_k)_{l}^\text{DE}-(m_k)_{14}^\text{GE}|}
  {\sum_k|(m_k)_{14}^\text{GE}|}, 
\end{equation}
which allows us to quantify the difference between the DE and the GE,
and
\begin{equation}\label{eq:bigdelta}
  \Delta(m^{\mathrm{ens}})_{l} = \frac{\sum_k|(m_k)_{l}^{\mathrm{ens}}-
    (m_k)_{14}^{\mathrm{ens}}|}{\sum_k|(m_k)_{14}^{\mathrm{ens}}|},  
\end{equation}
which allows us to estimate the convergence of the NLCE
calculations \cite{rigol_14a}. $\delta(K)_{l}$ and
$\Delta(K^{\mathrm{ens}})_{l}$ follow straightforwardly from
Eqs.~\eqref{eq:smalldelta} and \eqref{eq:bigdelta}, respectively, by
removing the sums and replacing $m_k\rightarrow K$.  For the GE
calculations when $T_I>1$, $(m_k)_{14}^\text{GE}$ and
$K_{14}^\text{GE}$ are converged within machine precision 
(see Ref.~\cite{supmat}).

Results for $\delta(m)_{l}$, $\delta(K)_{l}$,
$\Delta(m^{\mathrm{DE}})_{l}$, and $\Delta(K^{\mathrm{DE}})_{l}$ vs $l$
are reported in
Figs.~\ref{fig:deltaKinMom}(a)--\ref{fig:deltaKinMom}(d),
respectively, for six values of $h$. They show the following: (i) The
momentum distribution function (a nonlocal quantity) and the kinetic
energy (a local quantity) exhibit qualitatively similar
behavior. (ii) For $h\gtrsim3.5$, $\delta(m)_{l}$ and $\delta(K)_{l}$
do not change with increasing $l$, and are much larger than
$\Delta(m^{\mathrm{DE}})_{l}$ and $\Delta(K^{\mathrm{DE}})_{l}$, i.e.,
the former are expected to remain nonzero in the thermodynamic limit. 
This supports the existence of many-body localization in the thermodynamic
limit. (iii) For $h\lesssim3.0$, $\delta(m)_{l}$ and $\delta(K)_{l}$
decrease with increasing $l$, and are of the same order of magnitude as  
$\Delta(m^{\mathrm{DE}})_{l}$ and $\Delta(K^{\mathrm{DE}})_{l}$
(which also decrease with increasing system size).  Hence, the
differences between those observables in the DE and the GE are
expected to vanish in the thermodynamic limit, i.e., those values of
$h$ belong to the ergodic phase. In this phase, $\delta(m)_{l}$ and
$\delta(K)_{l}$ behave as in systems without disorder
\cite{rigol_14a}. (iv) $\Delta(m^{\mathrm{DE}})_{l}$ and
$\Delta(K^{\mathrm{DE}})_{l}$ in
Figs.~\ref{fig:deltaKinMom}(c) and \ref{fig:deltaKinMom}(d) show that the
NLCE convergence errors are largest in the region where the system
transitions between ergodic and localized.

\begin{figure}
  \includegraphics[width=0.45\textwidth]{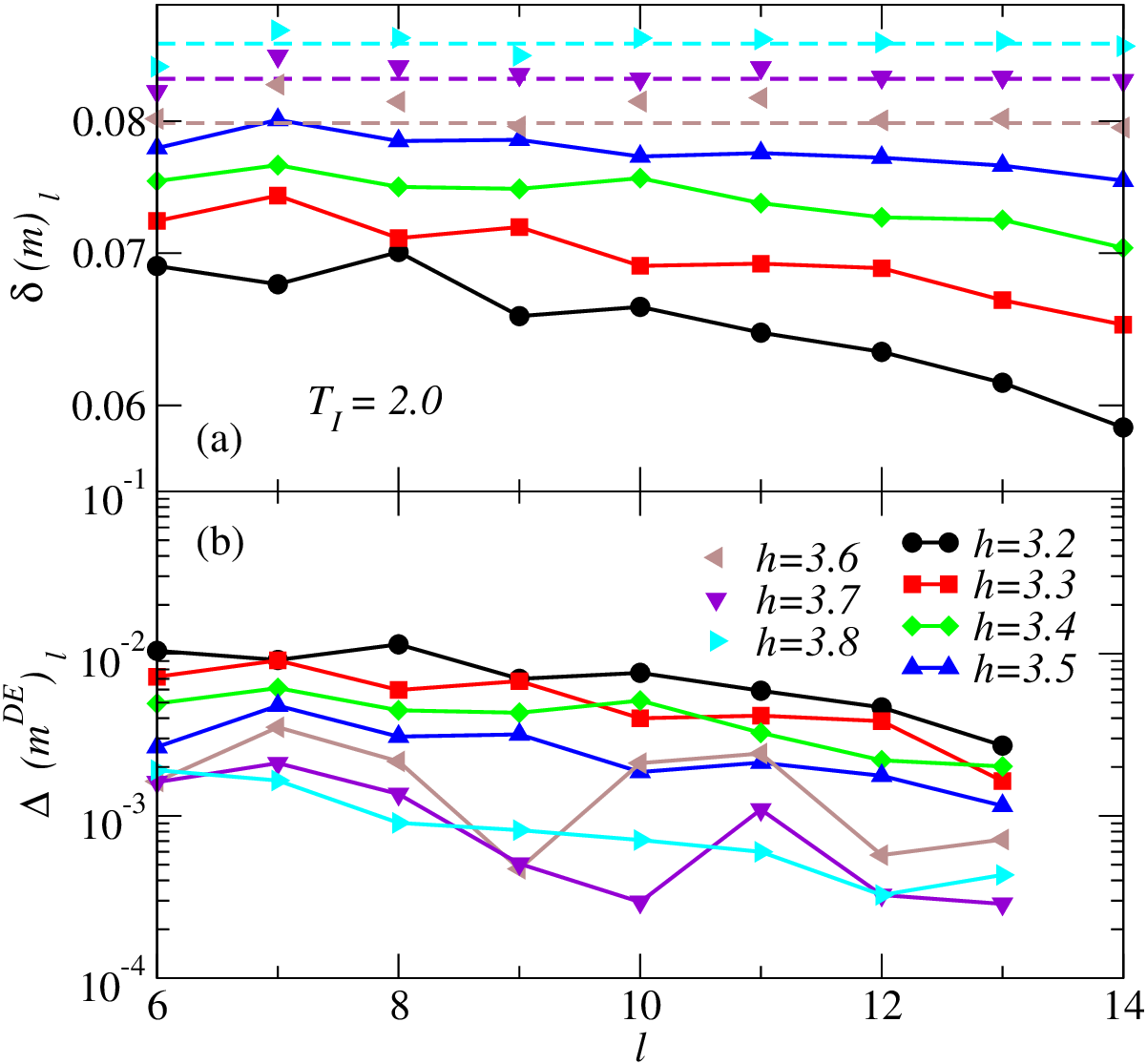}
  \caption{(Color online) Relative differences for the momentum
    distribution vs $l$ in the NLCE calculation for 
    $3.2 \leq h \leq 3.8$. (a) $\delta(m)_{l}$
    and (b) $\Delta(m^\text{DE})_{l}$. In (a), horizontal dashed lines
    correspond to the average value of last two orders of $\delta
    (m)_{l}$ for $h = 3.6$, $3.7$, and $3.8$.}
  \label{fig:deltaMom}
\end{figure}

In order to better pin down the transition point between the ergodic and
many-body localized phases, in Fig.~\ref{fig:deltaMom}(a) we plot
$\delta(m)_{l}$ vs $l$ in the vicinity of $h=3.5$.  For $h \geq 3.6$,
we see that $\delta(m)_{l}$ seems to saturate to a finite value that
is larger than $\Delta(m^\text{DE})_{13}$, suggesting that the system
is many-body localized for $h \geq 3.6$. The transition between
ergodic and many-body localized can occur for smaller values of $h$
as, for larger values of $l$, the plots for $\delta(m)_{l}$ may
saturate to a constant value. However, we expect that $h_c\approx 3.5$
since in the vicinity of this disorder strength we see that $\delta(m)_{l}$ 
and $\Delta(m^\text{DE})_{l-1}$ are very close to each other for the
largest system sizes studied. We should stress that, for $T_I> 2$,
we do not find indications that $h_c$ increases significantly with 
increasing $T_I$ \cite{supmat}. In general, it is expected that, as one 
increases the mean energy density after the quench (which is exactly 
what increasing $T_I$ does in our case), the transition point between
the delocalized and localized phases should move towards stronger 
disorder \cite{kjall14}. In the systems studied here, it is likely that 
a $T_I<2$ is needed to clearly observe that effect. However, the failure of NLCE
to converge in that regime does not allow us to check it.

In summary, we have studied quantum quenches in the thermodynamic
limit in an interacting model with binary disorder. This was possible
by generalizing the NLCE approach introduced in Ref.~\cite{rigol_14a} to
solve problems with disorder. We have shown that for quenches starting
in a delocalized phase, a freezing of correlations can occur in the 
steady state after the quench right after entering the many-body
localized phase. We located the critical value of the transition
between the ergodic and many-body localized phase using a quantum
chaos indicator (the average ratio between consecutive energy gaps) in
finite systems and the difference between NLCE predictions for
observables in the DE and the GE after quantum quenches. The values of
$h_c$ were found to be consistent in those two schemes. The small
convergence errors of NLCE for $h>h_c$ strongly support that the
many-body localized phase occurs in the thermodynamic limit. We 
should stress that the NLCE approach introduced here can be used 
to study disordered systems in equilibrium \footnote{B. Tang, D. Iyer, 
M. Rigol, arXiv:1501.00990.} and after quenches 
\footnote{B. Tang, D. Iyer, M. Rigol (unpublished).} 
in two (or higher) dimensions.

\paragraph*{Acknowledgments.} This work was supported by the Office of
Naval Research.

\bibliography{MBL}

\onecolumngrid

\vspace*{0.3cm}

\centerline{\large \bf Supplemental Material}

\vspace*{0.1cm}

\twocolumngrid

\subsection*{Convergence of NLCEs for the DE and the GE}

NLCEs, when converged, give exact results in the thermodynamic
limit. Here, we check the convergence of the calculations. We define the difference
\begin{equation}
  \label{eq:1}
  \Delta(\mathcal{O}^{\rm ens})_{l} \equiv \frac{|\mathcal{O}^{\rm ens}_{l}-\mathcal{O}^{\rm ens}_{14}|}
  {|\mathcal{O}^{\rm ens}_{14}|},
\end{equation}
where $\mathcal{O}$ is either the kinetic energy $K$ or the energy $E$. For the momentum, we define
\begin{equation}
  \Delta(m^{\mathrm{ens}})_{l} = \frac{\sum_k|(m_k)_{l}^{\mathrm{ens}}-
    (m_k)_{14}^{\mathrm{ens}}|}{\sum_k|(m_k)_{14}^{\mathrm{ens}}|},
\end{equation}
where $m_{k}$ is the momentum distribution function.
In all cases,
``ens'' refers to either the diagonal ensemble (DE) or the 
grand-canonical ensemble (GE). 

\begin{figure}[!t]
  \includegraphics[width=0.45\textwidth]{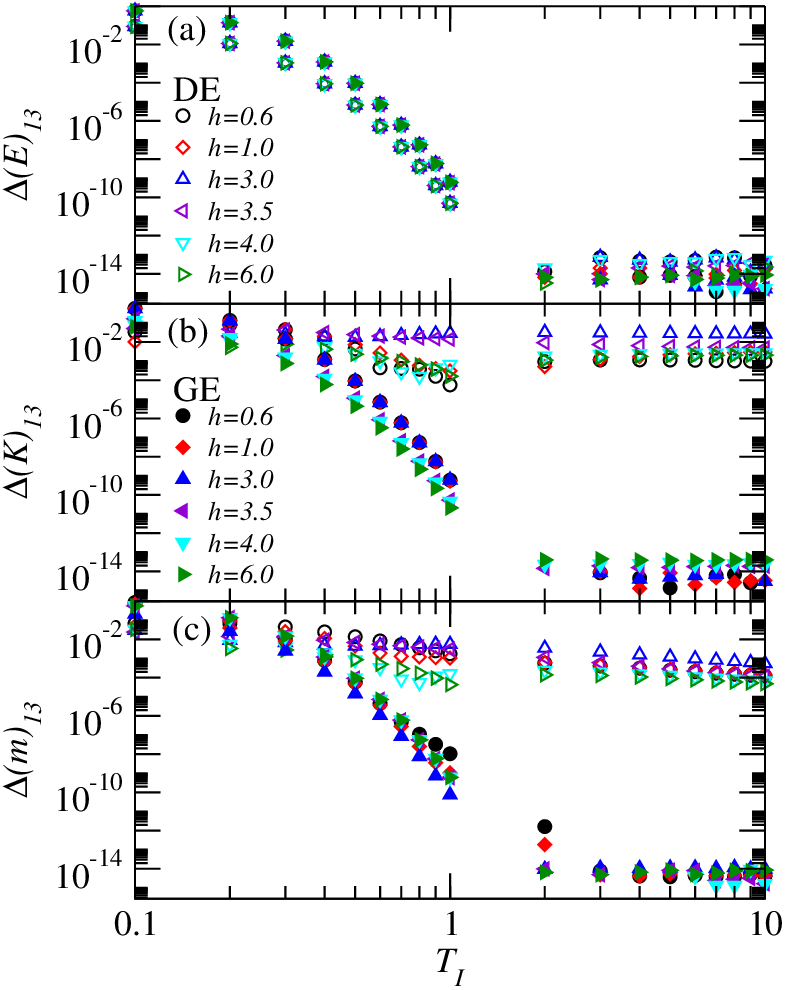}
  \vspace{-0.3cm}
  \caption{(Color online) Results for: (a) $\Delta(E^{\mathrm{ens}})_{13}$, 
  (b) $\Delta(K^{\mathrm{ens}})_{13}$, and (c) $\Delta(m^{\mathrm{ens}})_{13}$ 
  as a function of $T_{I}$ for the same set of quenches as in Fig.~3 in the main text.
  Open (filled) symbols depict the relative differences in the DE (GE). 
  In all panels, the GE results appear converged within machine precision 
  for temperatures $T_{I}\gtrsim 2$. For the DE, only the energy (a) converges
  within machine precision.}
  \label{fig:deltaVsT}
\end{figure}

In order to determine the initial temperature $T_{I}$ for which the
various observables calculated using NLCEs in the DE and GE are well
converged, we plot $\Delta(E^{\mathrm{ens}})_{13}$ in
Fig.~\ref{fig:deltaVsT}(a), $\Delta(K^{\mathrm{ens}})_{13}$ in
Fig.~\ref{fig:deltaVsT}(b), and $\Delta(m^{\mathrm{ens}})_{13}$ in
Fig.~\ref{fig:deltaVsT}(c) as a function of $T_{I}$ for the same set
of quenches as in Fig.~3 in the main text. Figure~\ref{fig:deltaVsT}
shows that, with increasing $T_{I}$, $\Delta(E^{\mathrm{GE}})_{13}$,
$\Delta(K^{\mathrm{GE}})_{13}$, $\Delta(m^{\mathrm{GE}})_{13}$
decrease and become zero within machine precison for $T_{I}>1.0$. 

We therefore expect that, within the cluster sizes accessible
to us, $E$, $K$, and $m$ in the GE have converged to the thermodynamic
limit results for $T_{I}> 1.0$.  In the DE, however, only the
energy [Fig.~\ref{fig:deltaVsT}(a)] coverges within machine precision.
As evident from Figs.~\ref{fig:deltaVsT}(b) and \ref{fig:deltaVsT}(c),
for the kinetic energy and the momentum distributions, respectively,
the relative errors do not change much with increasing temperature for
$T_{I} > 1.0$.  For these observables in the DE, the error can only be 
reduced by considering larger system sizes.

\begin{table}[!b]
\caption{Effective temperatures used in the GE calculations}
\begin{tabular}{| r | r | r | r | r | r | r |}
\hline\hline
  $T_{I}$  &   $h=0.6$   &     $h=1.0$  &    $h=3.0$  &    $h=3.5$  &     $h=4.0$  &  $h=6.0$ \\ \hline
    2.0    &   2.996 &    3.482 &   9.900 &  12.558 &   15.635 &   32.095  \\ \hline
   10.0    &  14.894 &   17.590 &  51.858 &  65.850 &   82.005 &  168.226  \\ \hline
   100.0   & 149.283 &  177.563 & 531.698 & 675.640 &  841.738 & 1727.657  \\ \hline \hline
\end{tabular}
\label{table:effectiveT}
\end{table}

\subsection*{Criticial disorder strength at higher temperature}

\begin{figure*}[!h]
  \includegraphics[width=0.45\textwidth]{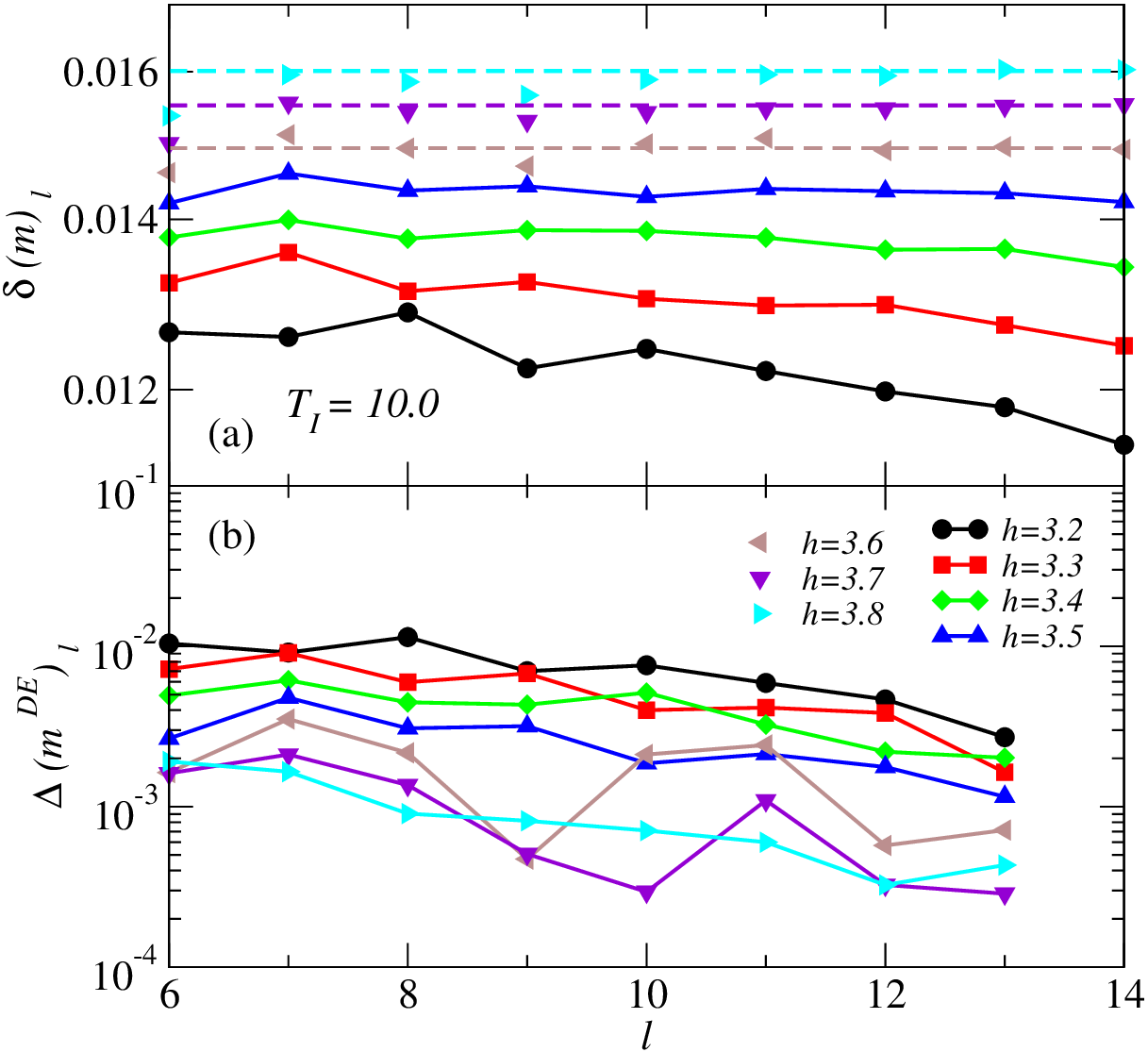}
    \includegraphics[width=0.45\textwidth]{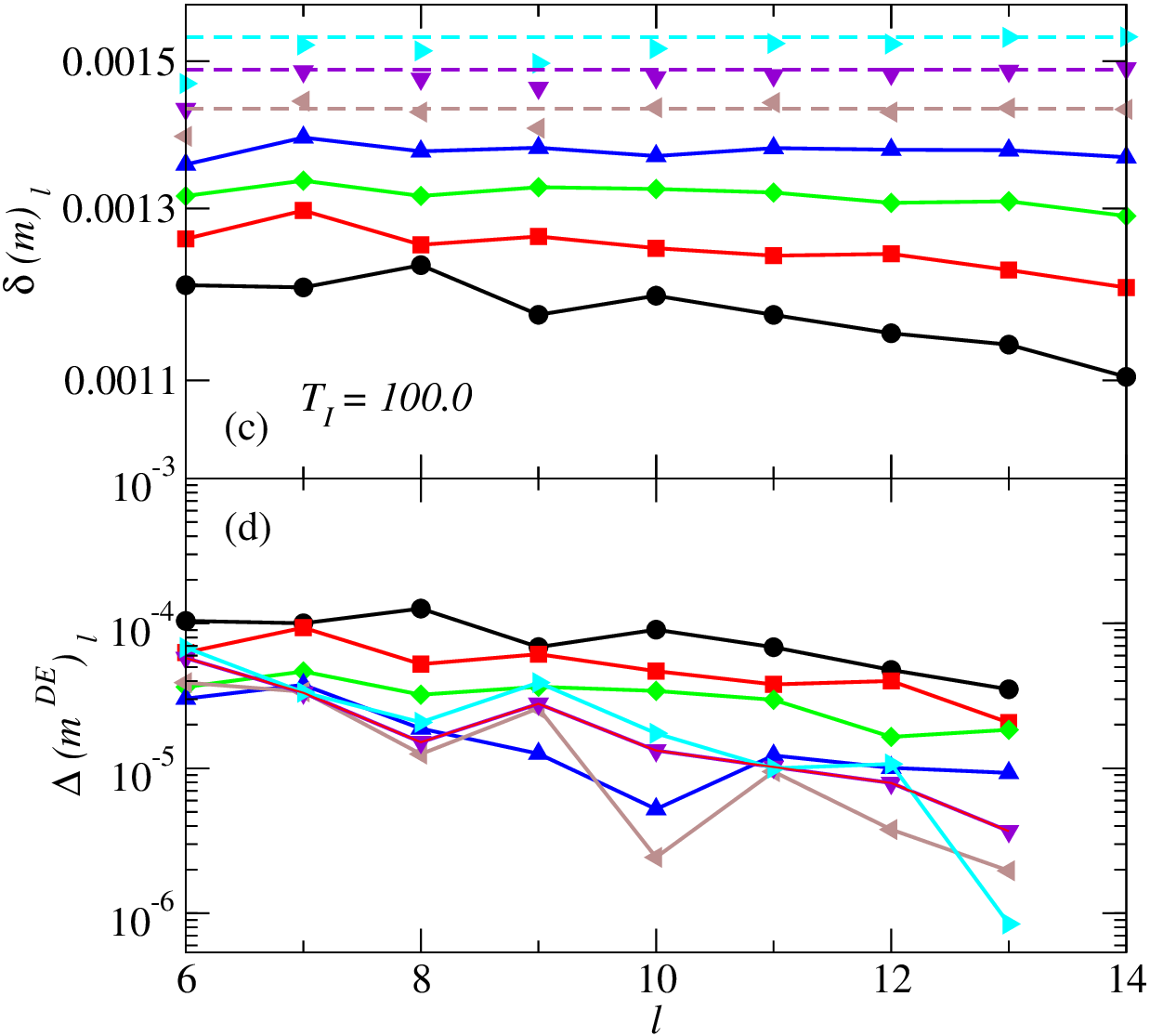}
    \vspace{-0.3cm}
  \caption{(Color online) The equivalent of Fig.~4 in the main text 
    for $T_{I}=10$ (left) and $T_{I}=100$ (right). For $h\geq 3.6$, 
    $\delta(m)_l$ vs $l$ [(a) and (c)] appears to converge to a nonzero 
    value with increasing system size. Furthermore, the convergence errors 
    [estimated by $\Delta(m^{\mathrm{DE}})_{l}$, see panels (b) and (d)] are 
    smaller than the $\delta(m)_l$ differences for those values of $h$. 
    These results are very similar to those for $T_{I}=2$ 
    reported in the main text.}
  \label{fig:deltaMomT10}
\end{figure*}

In Fig.~\ref{fig:deltaMomT10}, we show the equivalent of Fig.~4 
but for higher initial temperatures. As mentioned 
there, a higher temperature is expected to increase the value 
of the critical strength required for the localized phase to appear. 
However, for the temperatures at which our NLCEs for the energy converge
within machine precision, we do not observe any 
significant difference between the results for $T_{I}=2$, 10, and 100.
This is possibly because $T_{I}=2$ is already too high to see this effect.
The effective temperatures after the quench are reported in Table I.

\onecolumngrid

\end{document}